\documentclass[letter]{aa}
\usepackage{txfonts,astro,natbib,graphicx,amssymb}

\begin{document}
\def\sigt{\ensuremath{\sigma_T}}
\def\sigtdv{\ensuremath{\sigma_W}}
\def\tdust{\ensuremath{T_{\rm dust}}}
\def\ra{\ensuremath{\rightarrow}}

\title{CN in prestellar cores\thanks{Based on observations
carried out with the IRAM-30~m telescope. IRAM is supported
by INSU-CNRS/MPG/IGN.}}

\author{%
  P.~Hily-Blant\inst{1}\thanks{Currently at LAOG (UMR 5571) Observatoire de Grenoble}
  \and
  M.~Walmsley\inst{2}
  \and
  G.~Pineau~des~For\^ets\inst{3,4}
  \and
  D.~Flower\inst{5}
}

\offprints{P. Hily-Blant, \email{hilyblan@iram.fr}}

\institute{IRAM, Domaine Universitaire, 300 rue de la Piscine, 38406 Saint-Martin d'H\`eres, France
\and INAF, Osservatorio Astrofisico di Arcetri, Largo Enrico Fermi 5, I-50125 Firenze, Italy
\and IAS (UMR 8617 du CNRS), Universit\'{e} de Paris--Sud,  F-91405 Orsay, France
\and LERMA (UMR 8112 du CNRS), Observatoire de Paris, 61 Avenue de l'Observatoire, F-75014, Paris, France
\and Physics Department, The University, Durham DH1 3LE, UK}

\date{Received / Accepted}

\abstract{Determining the structure of and the velocity
field in prestellar cores is essential to understanding
protostellar evolution.}  {We have observed the dense
prestellar cores L~1544 and L~183 in the $N = 1 \rightarrow
0$ rotational transition of CN and \thcn \ in order to test
whether CN is depleted in the high--density nuclei of these
cores.}  {We have used the IRAM~30~m telescope to observe
along the major and minor axes of these cores. We compare
these observations with the 1~mm dust emission, which serves
as a proxy for the hydrogen column density.}{ We find that
while CN\jone\ is optically thick, the distribution of
\thcn\jone\ intensity follows the dust emission well,
implying that the CN abundance does not vary greatly with
density.  We derive an abundance ratio of $\rm
[CN]/[\hh]=\dix{-9}$ in L~183 and $1-3$\tdix{-9} in L~1544,
which, in the case of L~183, is similar to previous
estimates obtained by sampling lower--density regions of the
core.}{We conclude that CN is not depleted towards the
high--density peaks of these cores and thus behaves like the
N-containing molecules \nnhp\ and \nhhh. CN is, to our
knowledge, the first C--containing molecule to exhibit this
characteristic.}


\authorrunning{Hily-Blant \etal}

\titlerunning{CN in prestellar cores}

\maketitle

\section{Introduction}

Knowledge of the structure and kinematics of prestellar
cores is important to our understanding of protostellar
evolution.  Cores which are close to the critical point at
which collapse sets in are representative of the preliminary
phase of evolution which subsequently leads to the formation
of a protostar.  In this context, the cores with the highest
central density and/or column density are the most
interesting; but the fact that CO and many other tracers of
the kinematics deplete on to dust grain surfaces at
densities above \dix{5}~\ccc\ \citep{tafalla2002} has been an
obstacle to progress in this area.  Fortunately,
observations have suggested that some N--containing species,
such as \nhhh\ and \nnhp, remain in the gas phase at
densities above \dix{5}~\ccc. Given that both \nhhh\ and
\nnhp\ form from \nn\, and that the volatility of \nn\ is
similar to that of CO \citep{oberg2005}, it is puzzling that
even these species survive. There is some evidence that they
finally freeze out at densities $\gtrsim\dix{6}$~\ccc\
\citep{bergin2002,belloche2002,pagani2007}, and,
irrespective of the reason for this behaviour, it would be
useful to find other tracers of the densest gas in the
cores.

Radicals represent a type of ``molecule'' which tends to
resist depletion; this comes about because, typically, they
are both formed and destroyed by neutral--neutral
substitution reactions in which both formation and
destruction depend in the same manner on the abundance of an
atom, such as C, N, or O. An example is NO, which is
believed to be formed and destroyed by reactions with atomic
nitrogen. Under these conditions, the fractional abundance
of NO is independent of the degree of depletion of atomic
N. In a previous article \citep[][\, hereafter
A07]{akyilmaz2007}, we attempted to test this hypothesis by
observing NO in two high density cores. We found, to our
discomfiture, that the hypothesis was incorrect: NO becomes
depleted in the high density cores of L~1544 and L~183
similarly to many other species.  The explanation of this
observational result is not entirely clear, but it probably
has to do with the fraction of oxygen which remains in the
gas phase at high densities.

The motivation for our NO study arose partly from the issues
of the nitrogen chemistry, referred to earlier. NO has been
considered to be the main intermediary between atomic and
molecular nitrogen, which are probably the main forms of
elemental nitrogen in the gas phase in molecular clouds.
However, the fact that NO appears to be not present in the
densest gas suggested that there may be other intermediaries
between N and \nn\ and that a possible candidate might be
CN. CN can form from hydrocarbons, such as CH, in reactions
with atomic N \citep{pineau1990} and is destroyed, also by
N, forming \nn.  Thus, if elemental carbon is present in the
dense gas, CN might mediate in the formation of \nn\ and
hence in the formation of \nhhh\ and \nnhp. However, the
fraction of elemental carbon which is available to form
hydrocarbons in the dense regions of prestellar cores is
uncertain. The reason for this uncertainty is that the main
reservoir of carbon in the gas phase of molecular clouds is
CO, and if CO freezes out, it might be expected that there
remains little carbon to form hydrocarbons and, by
extension, CN.  However, from an observational point of
view, one knows only that CO is at least an order of
magnitude underabundant in prestellar core nuclei. There are
non-thermal processes \citep[\eg][]{leger1985} which can
maintain gas phase CO at abundances of order $\rm [CO]/[\hh]
= 10^{-6}$ at densities of order \dix{6}~\ccc\ and which
might suffice to produce appreciable amounts of \cp, a
precursor of CH, following the reaction of CO with
cosmic--ray produced \hep.

The present Letter presents observations of CN and \thcn\
which demonstrate that CN is relatively abundant in the high
density cores of L~183 and L~1544.  In section~2, we
summarize the observations and, in section~3, we present the
basic observational results. Section~4 contains a brief
discussion of the implications of our observational results
for chemical models.

\section{$^{12}$CN and $^{13}$CN\jone\ observations}

\begin{figure}
  \def\wa{0.5\hsize}
  \begin{center}
    \includegraphics[width=\wa,angle=0]{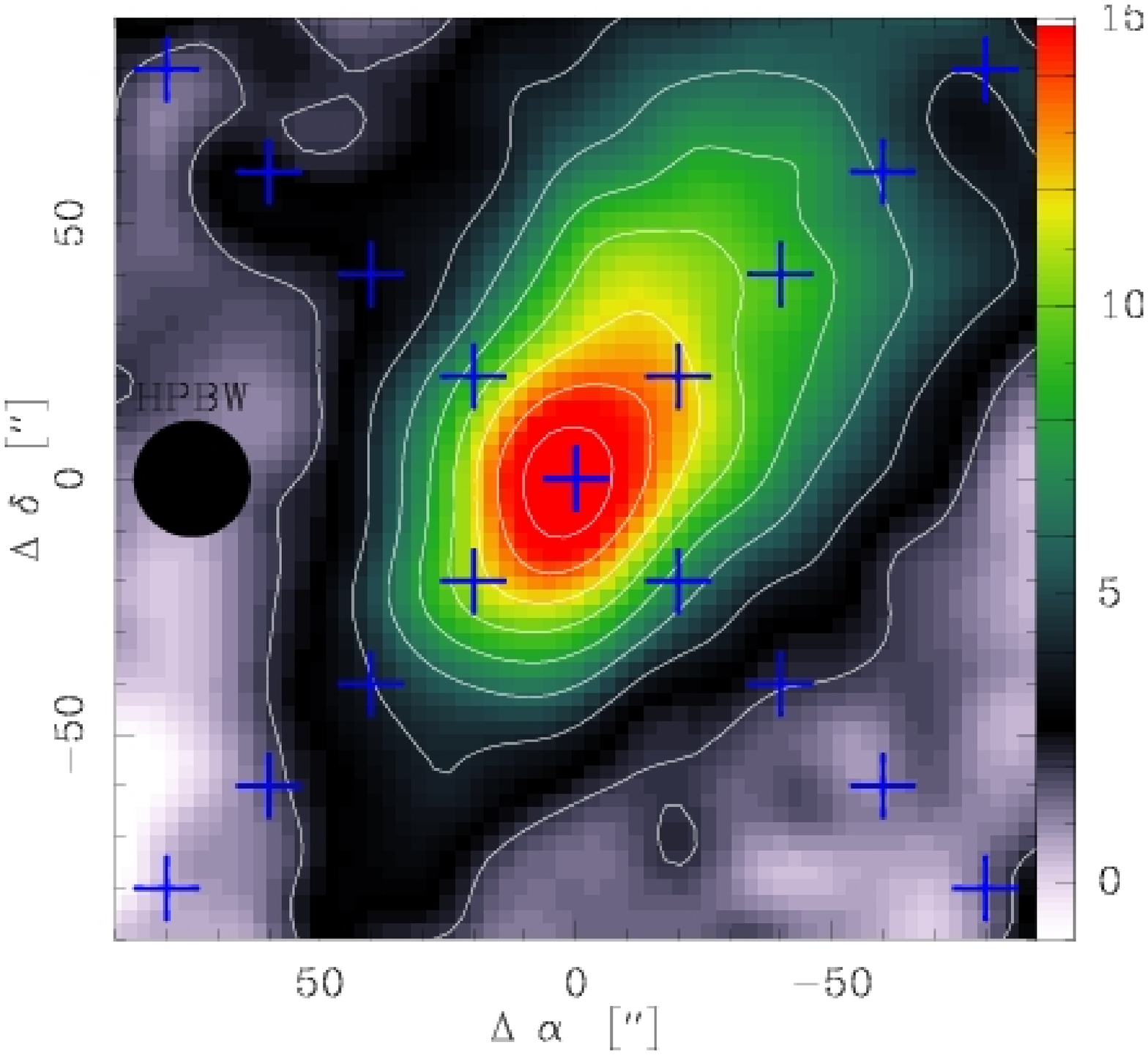}\hfill%
    \includegraphics[width=\wa,angle=0]{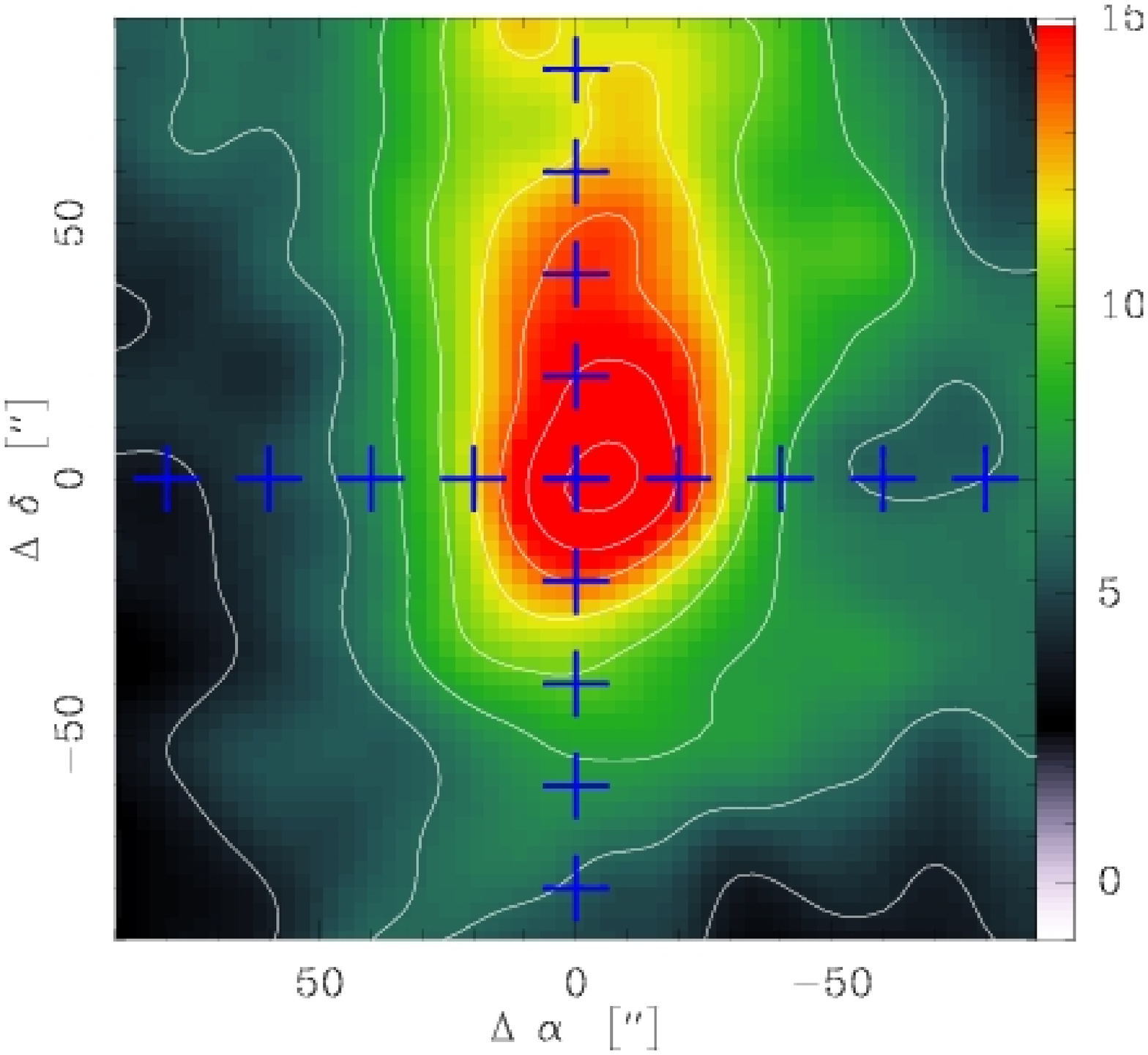}
    \caption{Dust emission at 1.3~mm (in \mjysr) towards
      L~1544 (left) and L~183 (right) from \cite{ward1999}
      and \cite{pagani2004}, respectively. The crosses
      indicate the positions where the \thcn\jone\ emission
      was measured (North is to the top). In L~1544, the
      NE--SW and NW--SE cuts are referred to as the minor
      and major axes, respectively. In L~183, major and minor
      refer to N--S and E--W, respectively.  The 22.8\arcsec\
      HPBW at 108.8~GHz is indicated. The $(0,0)$ position
      corresponds to $\alpha(2000)=\hms{05}{04}{16.9},
      \delta(2000)=\dms{25}{10}{47}$ for L~1544 and
      $\alpha(2000)=\hms{15}{54}{08.8},
      \delta(2000)=\dms{-02}{52}{44}$ for L~183.}
    \label{fig:dust}
  \end{center}
\end{figure}

Observations were carried out with the IRAM 30~m
telescope. The CN$(N = 1\rightarrow 0, F_2 = 3/2\rightarrow
1/2)$ multiplet (113.5~GHz: see Table~\ref{tab:ratios12cn})
was observed at different epochs between December 2006 and
August 2007, under average to good weather
conditions. Observations of the \thcn$(N = 1\rightarrow 0,
F_2 = 2\rightarrow 1)$ multiplet \citep[at
108.8~GHz][]{gerin1984} were performed in August 2007 under
good conditions. The HPBW at the CN and \thcn\ frequencies
are 22 and 22.8\arcsec, respectively. The VESPA facility
autocorrelator was used to obtain 20~kHz channel spacing,
with 60~MHz and 35~MHz bandwidths for CN and \thcn,
respectively. Hence we cover the five strongest hyperfine
structure (HFS) transitions of both isotopologues. The
observing strategy was identical for all measurements:
frequency--switching mode with a 7.8~MHz throw and a phase
time of 0.5~s, with atmospheric and bandpass amplitude
calibrations every 10 to 15~min. Data reduction and analysis
were done with the CLASS software \citep{IRAM_report_2005-1}
of the GILDAS facility \citep{pety_gildas}. Instrumental
bandpass and atmospheric contributions were subtracted with
polynomial baselines, before and after the folding of the
two--phase spectra. Concerning the CN spectra, the SSB
receiver and system temperatures were $\trec=75$~K and
$\tsys\approx170$~K, with zenith opacity 0.13. The final
rms, in the main--beam temperature (\tmb), in each channel
of width $\delta v=0.052$~\kms\ is $\sigt\approx 70$~mK for
both L~1544 and L~183. Regarding \thcn, $\trec=70-90$~K and
$\tsys=100-250$~K, resulting in signal band zenith opacities
$0.05-0.1$. The final rms in each channel of width $\delta
v=0.054$~\kms\ are $\sigt\approx 15$~mK towards L~1544 and
17~mK towards L~183. In what follows, all temperatures are
on the main--beam scale, $\tmb=\feff\tant/\beff$ where
\tant\ is the antenna temperature corrected for atmospheric
absorption, and the forward and beam efficiencies are
respectively $\feff=0.95$ and $\beff=0.75$.

The observations were made along two cuts, following the
major and minor axes of L~1544 and L~183: see
Fig.~\ref{fig:dust}.

\section{Results}

\subsection{$^{12}$CN}

\begin{figure}
  \def\wa{0.6\hsize}
  \begin{center}
    \includegraphics[width=0.8\hsize,angle=-90]{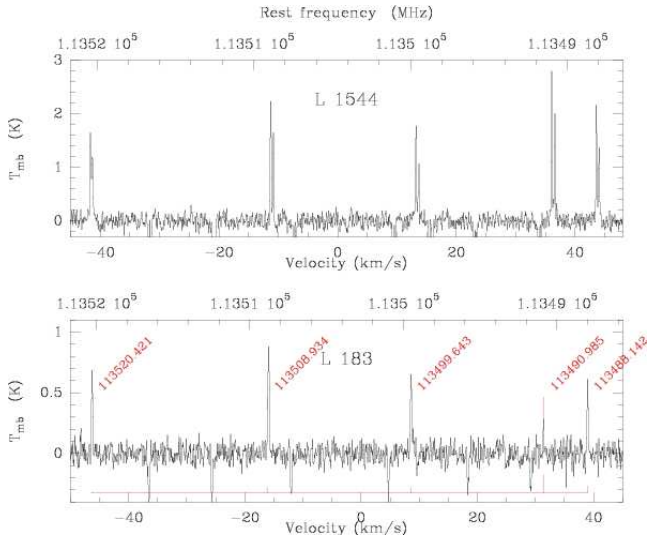}
    \caption{CN$(N = 1\ra 0, F_2 = 3/2\ra 1/2)$ spectrum
      towards the $(0,0)$ positions in L~1544 (top) and
      L~183 (bottom). The relative line intensities expected
      under optically thin LTE conditions are shown towards
      the bottom of the lower panel.}
    \label{fig:cn10zero}
  \end{center}
\end{figure}

\begin{figure}
  \def\wa{0.4\hsize}
  \begin{center}
    \includegraphics[width=\wa,angle=-90]{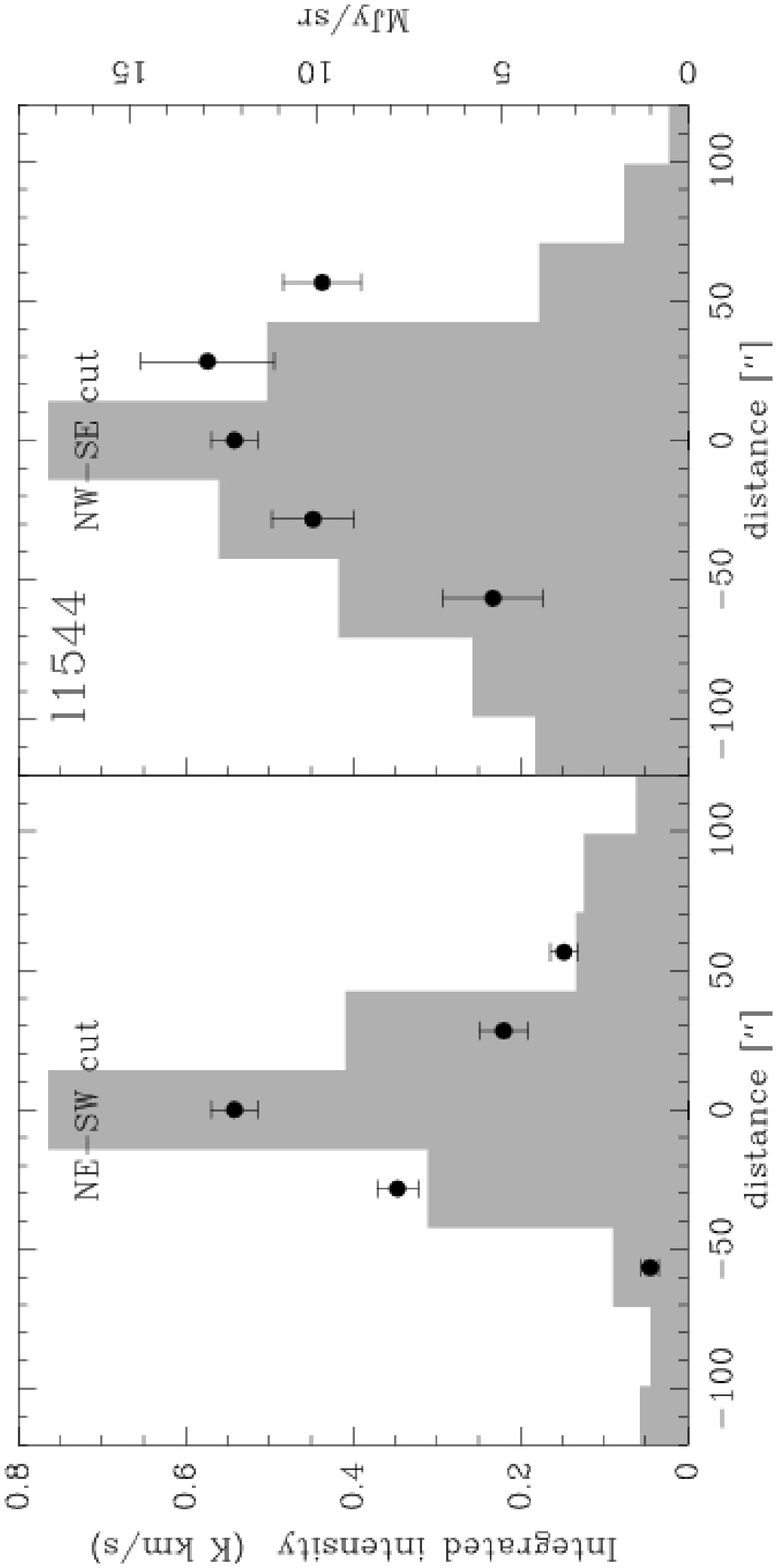}\\
    \includegraphics[width=\wa,angle=-90]{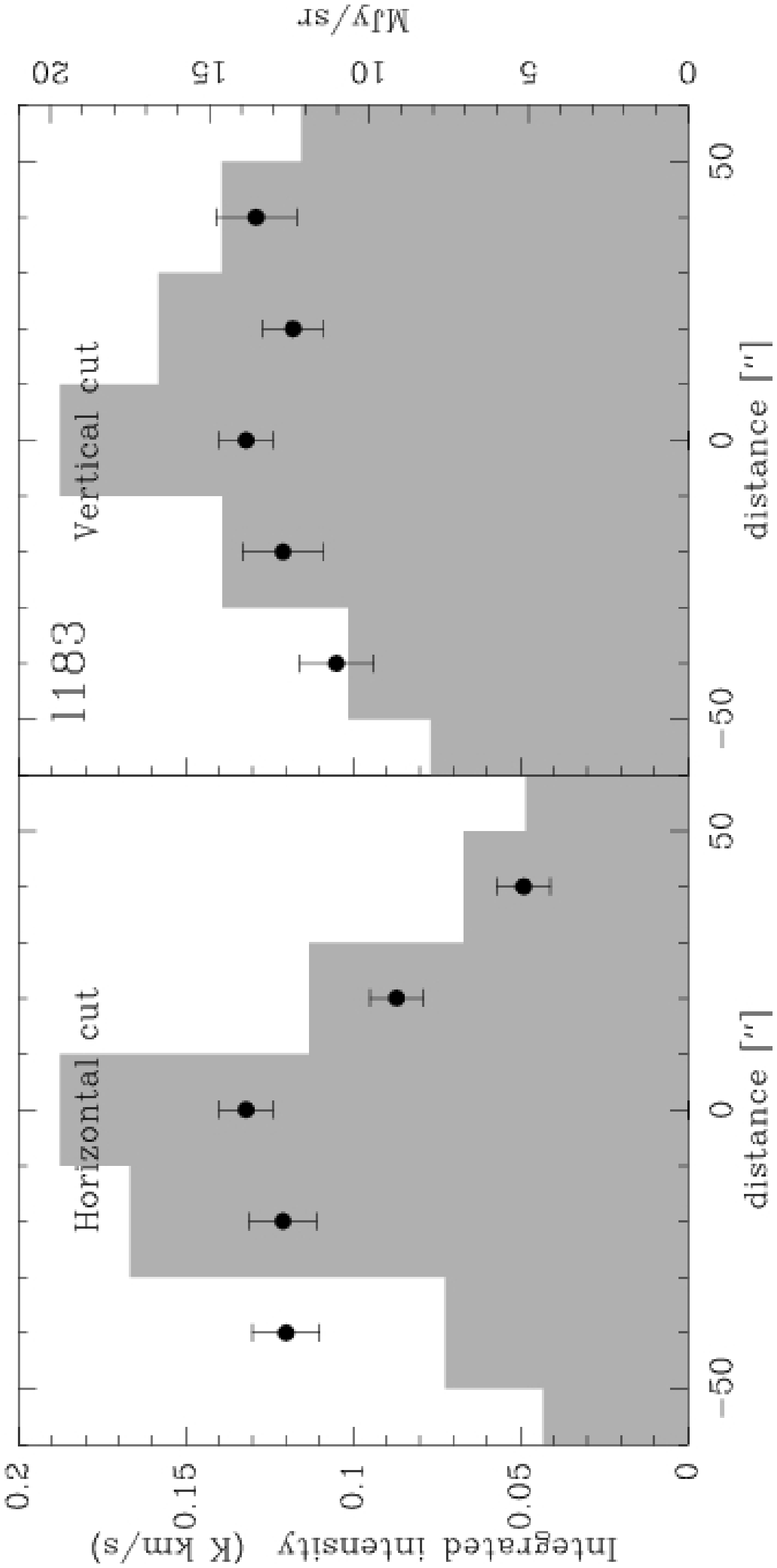}
    \caption{Cuts of the integrated intensity (\kkms) (left hand scale)
      of the weakest HFS component of
      CN$(N = 1\ra 0, F_2 = 3/2\ra 1/2)$, at rest
      frequency 113520.4~MHz and with relative intensity
      0.0123, towards L~1544 (top) and L~183 (bottom). The light
      histogram is the dust emission, in MJy\,\unit{sr}{-1} (right
      hand scale). }
    \label{fig:cn10cut}
  \end{center}
\end{figure}

In Fig.~\ref{fig:cn10zero}, we show our CN$(N = 1\rightarrow
0, F_2 = 3/2\rightarrow 1/2)$ spectra towards the $(0,0)$
positions in L~183 and L~1544. It is noteworthy that the
L~1544 profiles are double--peaked (see
Fig.~\ref{fig:profile_l1544}) and, more significantly, that
the observed intensity ratios of the HFS components in both
sources are inconsistent with the values expected under
optically thin LTE conditions; we list the expected and
observed ratios in Table~\ref{tab:ratios12cn}.  In L~183,
the transition of highest intrinsic strength, at
113490.985~MHz, is far from being the most intense line
observed. The most likely explanation is that the lines are
optically thick. Accordingly, we decided to observe the
corresponding \thcn\ lines. It is striking that the
$^{12}$CN lines peak in roughly the same position as the
dust millimeter emission; this is apparent in
Fig.~\ref{fig:cn10cut}, where we plot the integrated
intensity of the (intrinsically weakest and hence most
likely to be optically thin) $F\rightarrow F^{'}=
1/2\rightarrow 3/2$ line at 113520.34~MHz against offset
from (0,0).  One sees that, at least in L~1544, the line
intensity is highest not far from the peak in the dust
emission, suggesting that the CN column density is a maximum
at that position.

\subsection{\th CN}

\begin{figure}
  \def\wa{0.55\hsize}
  \begin{center}
    \includegraphics[width=0.8\hsize,angle=-90]{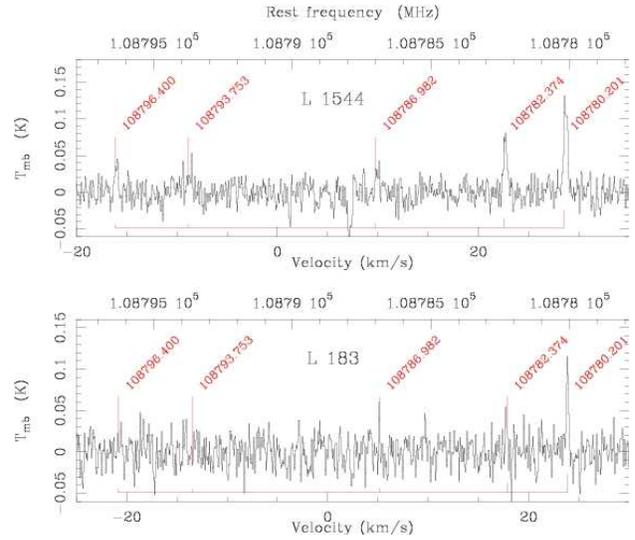}
    \caption{\thcn$(N = 1\ra 0, F_2 = 2\ra 1)$ spectrum at positions
      $(0,0)$ in L~1544 (top) and L~183
      (bottom). The five observed HFS components are
      indicated.}
    \label{fig:zero}
  \end{center}
\end{figure}

Figure~\ref{fig:zero} shows the (0,0) \thcn$(N =
1\rightarrow 0, F_2 = 2\rightarrow 1)$ spectrum towards the
two cores. In both cores, the emission in the main
$F=3\rightarrow 2$ 108780.201~MHz component was successfully
detected at all offsets smaller than 40\arcsec. Towards
L~183, we detect only the main HFS component at most
positions, although the $F = 2\rightarrow 1$ transition at
108782.37~MHz is seen also at offset $(20,0)$. Towards
L~1544, we detect several components and can determine their
relative intensities.  At each position, a Gaussian profile
was fitted to the main component (with relative intensity
$R=0.194$) from which we obtain the integrated intensity,
$W_{\rm main}$ (on the \tmb\ scale). At some positions
toward L~1544, the lines are double--peaked
\citep{williams1999,vdtak2005}, in which case a
two--component Gaussian fit was made. The line parameters
are given in Tables~\ref{tab:l1544} and \ref{tab:l183}.

We note that the intensity ratios of the HFS components are
consistent with low optical depths, which we have assumed
when estimating the CN column density. We assume also that
the level populations are consistent with LTE at a
temperature of 8~K \citep[a compromise between estimates
quoted in the literature, \eg][]{crapsi2007,pagani2007}. For
the [\tw CN]/[\thcn] abundance ratio, we adopt a value of 68
\citep{milam2005}.

\begin{figure}
  \def\wa{0.43\hsize}
  \begin{center}
    \includegraphics[width=\wa,angle=-90]{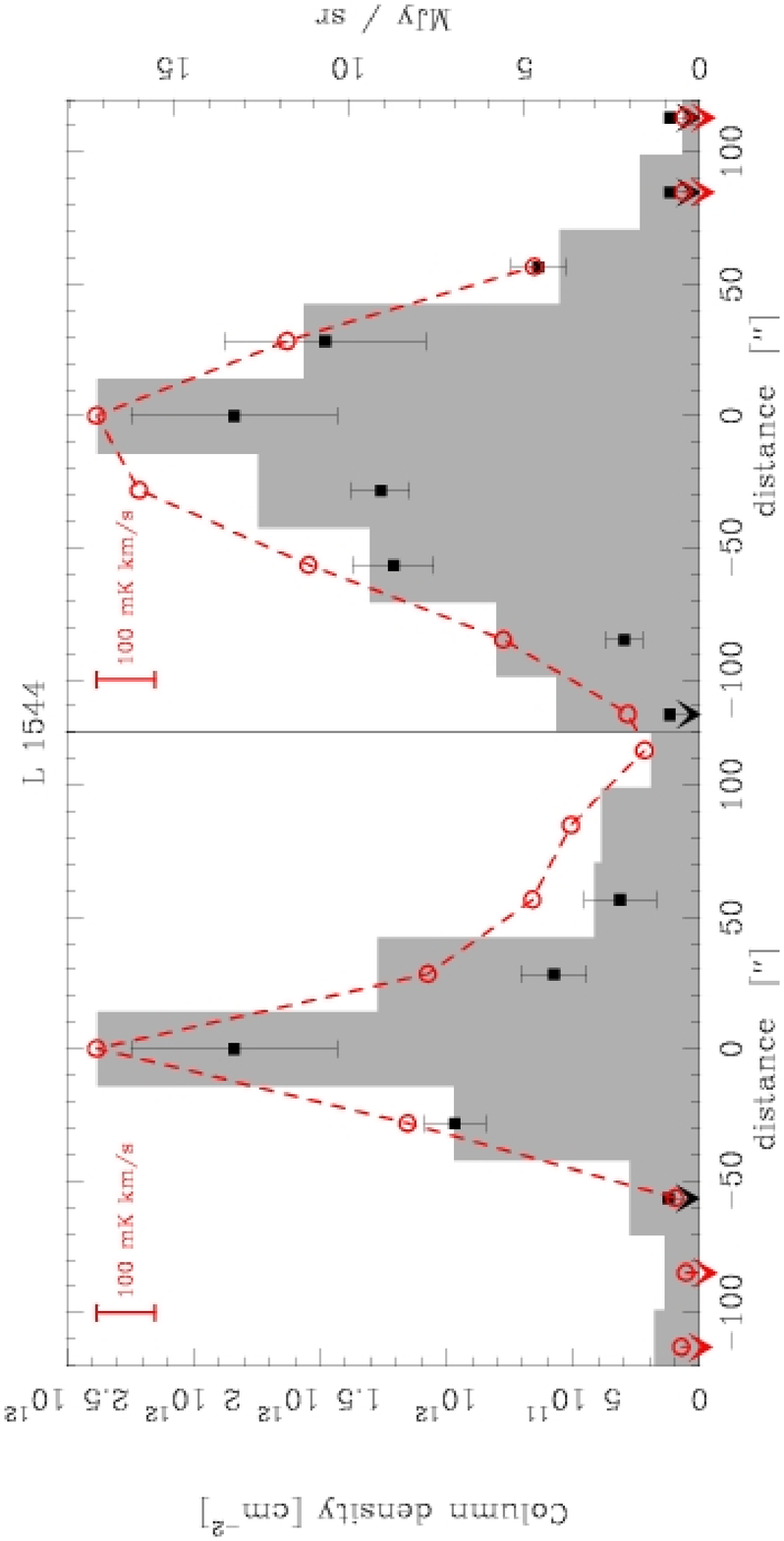}\\
    \includegraphics[width=\wa,angle=-90]{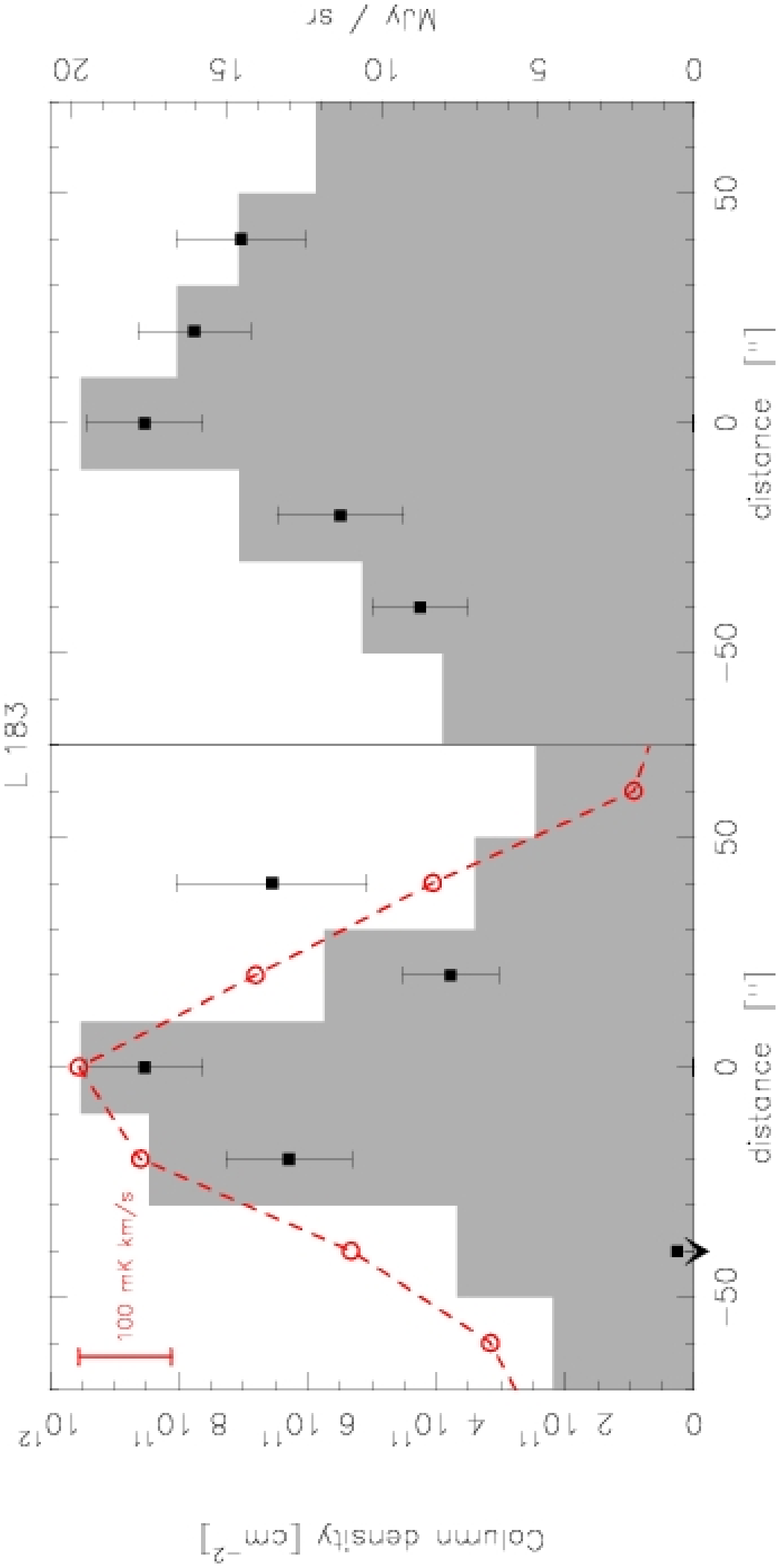}
    \caption{Cuts through L~1544 (top) and L~183
      (bottom). Filled squares: \thcn total column
      density (Tables~\ref{tab:l1544} and ~\ref{tab:l183})
      in \cc\ (left hand scale). Open red circles:
      integrated intensity (in m\kkms: scale is indicated in
      top left corner) of the $(J,F_1,F)=1,0,1 \ra 0,1,2$
      transition of \nnhp\ (see A07). Light histogram: dust
      emission in MJy\,\unit{sr}{-1} (right hand scale).}
    \label{fig:cut}
  \end{center}
\end{figure}

Subject to the above assumptions, we derive the column
density of \thcn\ from the integrated intensity of the
strongest hyperfine component at 108780.201~MHz, which was
taken to have a relative intensity $R=0.194$ of the total
\citep[][\, and Table~\ref{tab:spectro}]{bogey1984}. We used
the formulation of A07.  Our results are summarized in
Tables~\ref{tab:l1544} and \ref{tab:l183} and shown in
Fig.~\ref{fig:cut}, where the error bars on the integrated
intensity are $\sigt \sqrt{{\rm FWHM} \times \delta v}$.

From the continuum intensity at 1.3~mm, smoothed to the
resolution of the \thcn\jone\ beam, we computed the \hh\
column density as $N(\hh) = S_{1.3\rm mm} / (\kappa\mu
m_{\rm H}B_{\nu}(\tdust))$, where $\mu=2.33$ is the mean
molecular weight of the gas, $m_{\rm H}$ is the mass of
atomic hydrogen and $\kappa=0.01$~cm$^2$~g$^{-1}$ is the
dust opacity per unit mass column density
\citep{genzel1992}. The error bars are computed from the
dispersion of the brightness inside the 22\arcsec\ beam. The
derived fractional abundances of \thcn, with respect to \hh
and at the dust peak, are 2.8\tdix{-11} and 1.3\tdix{-11} in
L~1544 and L~183, respectively. These convert into CN
fractional abundances of 1.9\tdix{-9} and 0.8\tdix{-9},
assuming $\rm [\tw CN]/[\thcn]=68$ \citep{milam2005}.

From Fig.~\ref{fig:cut}, we conclude that \thcn \ follows
the dust and hence that the CN fractional abundance does not
vary greatly with density in the centres of these cores;
this is in sharp contrast to results for other species, in
particular to observations of CO \citep{caselli1999} and NO
(A07). It is also in contrast to the results of
\cite{tafalla2006}, who found that, in L~1498 and L~1517B,
most C--containing molecules (though CN was not observed)
deplete towards the density peaks in these objects. To our
knowledge, CN is the first C--containing species which
appears to avoid ``depletion'' towards the density peaks of
prestellar cores, although more observations are required to
confirm this general statement.  The CN fractional
abundances which we derive vary little with offset (and, by
inference, with density), as Tables ~\ref{tab:l1544} and
~\ref{tab:l183} show. In the case of L~183, they differ
little also from the CN fractional abundances derived from
observations with a larger beam and at various positions in
the core \citep{dickens2000}.  We conclude that, even in
situations where CO (the main sink of C) is highly depleted,
some C--containing species can survive to densities of the
order of to $106$~\ccc.

\def\taumain{\ensuremath{\tau_{\rm m}}}
\def\cdust{\ensuremath{N_{\rm dust}}}
\begin{table}
  \begin{center}
    \caption{\thcn\ column density in L~1544, $N(\thcn)$ (in
      units of \dix{12}\,\cc), with $W = W_{\rm main}/R$ (in
      m\kkms) and $R=0.194$ (see Table~\ref{tab:spectro}). A
      single excitation temperature was adopted,
      $\texc=8$~K. The \hh\ column density is computed
      assuming $\tdust=\texc$. The CN fractional abundance
      $X({\rm CN})=[\rm CN]/[\hh]$ was computed assuming
      $\rm [\tw CN]/[\thcn]=68$ \citep{milam2005}. Error
      bars are 1$\sigma$.}
    \scriptsize
    \begin{tabular}{c @{,} c c r r r r}\hline\hline
      $\delta x$ & $\delta y$ & $W$ &  $N(\thcn)$ & $S_{\rm 1.3mm}$ & $N(\hh)$ & $X({\rm CN})$\\
      \arcsec & \arcsec & m\kkms & \dix{12}\cc & MJy \unit{sr}{-1} & \dix{22} \cc & \dix{-9}\\\hline
     -80 &      80 &        $\le$ 25 &      $\le$ 0.1 &     4.1$\pm$0.4 &   1.6$\pm$ 0.1 &     $\le$0.5\\
     -60 &      60 &      60$\pm$ 15 &   0.3$\pm$ 0.1 &     5.8$\pm$0.6 &   2.3$\pm$ 0.2 &  0.9$\pm$0.2\\
     -40 &      40 &     235$\pm$ 30 &   1.2$\pm$ 0.2 &     9.4$\pm$0.4 &   3.6$\pm$ 0.2 &  2.3$\pm$0.3\\
     -20 &      20 &     245$\pm$ 20 &   1.3$\pm$ 0.1 &    12.6$\pm$0.8 &   4.9$\pm$ 0.3 &  1.7$\pm$0.2\\
       0 &       0 &     360$\pm$ 80 &   1.8$\pm$ 0.4 &    17.2$\pm$0.5 &   6.6$\pm$ 0.2 &  1.9$\pm$0.4\\
      20 &     -20 &     290$\pm$ 75 &   1.5$\pm$ 0.4 &    11.3$\pm$1.1 &   4.4$\pm$ 0.5 &  2.3$\pm$0.7\\
      40 &     -40 &     125$\pm$ 20 &   0.6$\pm$ 0.1 &     4.0$\pm$0.6 &   1.6$\pm$ 0.2 &  2.7$\pm$0.6\\
      60 &     -60 &        $\le$ 25 &      $\le$ 0.1 &     1.7$\pm$0.5 &   0.7$\pm$ 0.2 &     $\le$1.2\\
      80 &     -80 &        $\le$ 25 &      $\le$ 0.1 &     0.5$\pm$0.4 &   0.1$\pm$ 0.1 &     $\le$5.8\\
      \hline
     -40 &     -40 &        $\le$ 25 &      $\le$ 0.1 &     2.0$\pm$0.3 &   0.8$\pm$ 0.1 &     $\le$1.0\\
     -20 &     -20 &     190$\pm$ 25 &   1.0$\pm$ 0.1 &     7.0$\pm$1.4 &   2.8$\pm$ 0.5 &  2.3$\pm$0.5\\
       0 &       0 &     360$\pm$ 80 &   1.8$\pm$ 0.4 &    17.2$\pm$0.5 &   6.6$\pm$ 0.2 &  1.9$\pm$0.4\\
      20 &      20 &     110$\pm$ 25 &   0.6$\pm$ 0.1 &     9.2$\pm$1.6 &   3.6$\pm$ 0.6 &  1.1$\pm$0.3\\
      40 &      40 &      60$\pm$ 25 &   0.3$\pm$ 0.1 &     3.0$\pm$0.3 &   1.2$\pm$ 0.1 &  1.8$\pm$0.8\\
      \hline
    \end{tabular}
    \label{tab:l1544}
  \end{center}
\end{table}

\begin{table}
  \begin{center}
    \caption{\thcn\jone\ column density in L~183 (see
      Fig.~\ref{fig:cut}). As Table~\ref{tab:l1544},
      with the same excitation temperature, $\texc=8$~K.}
    \scriptsize
    \begin{tabular}{c @{,} c c r r r r}\hline\hline
      $\delta x$ & $\delta y$ & $W$ & $N(\thcn)$ & $S_{\rm
      1.3mm}$ & $N(\hh)$ & $X({\rm CN})$\\ \arcsec &\arcsec
      & m\kkms& \dix{12}\cc & MJy \unit{sr}{-1} & \dix{22}
      \cc & \dix{-9}\\\hline
     -40&0 &        $\le$  5 &      $\le$ 0.0 &  8.2$\pm$1.1 &   3.3$\pm$ 0.4 &     $\le$0.1\\
     -20&0 &     125$\pm$ 20 &   0.6$\pm$ 0.1 & 15.8$\pm$1.4 &   6.1$\pm$ 0.6 &  0.7$\pm$0.1\\
       0&0 &     165$\pm$ 20 &   0.9$\pm$ 0.1 & 18.3$\pm$0.5 &   7.0$\pm$ 0.2 &  0.8$\pm$0.1\\
      20&0 &      75$\pm$ 15 &   0.4$\pm$ 0.1 & 12.2$\pm$1.4 &   4.8$\pm$ 0.6 &  0.5$\pm$0.1\\
      40&0 &     130$\pm$ 30 &   0.7$\pm$ 0.1 &  7.1$\pm$0.7 &   2.8$\pm$ 0.3 &  1.6$\pm$0.4\\
      \hline
       0&-40 &      85$\pm$ 15 &   0.4$\pm$ 0.1 &  9.8$\pm$0.6 &   3.9$\pm$ 0.2 &  0.7$\pm$0.1\\
       0&-20 &     105$\pm$ 20 &   0.5$\pm$ 0.1 & 14.2$\pm$1.2 &   5.6$\pm$ 0.5 &  0.7$\pm$0.1\\
       0&  0 &     165$\pm$ 20 &   0.9$\pm$ 0.1 & 18.3$\pm$0.5 &   7.0$\pm$ 0.2 &  0.8$\pm$0.1\\
       0& 20 &     150$\pm$ 15 &   0.8$\pm$ 0.1 & 16.1$\pm$0.6 &   6.3$\pm$ 0.2 &  0.8$\pm$0.1\\
       0& 40 &     135$\pm$ 20 &   0.7$\pm$ 0.1 & 14.5$\pm$0.2 &   5.7$\pm$ 0.1 &  0.8$\pm$0.1\\
       \hline
    \end{tabular}
    \label{tab:l183}
  \end{center}
\end{table}


\section{Conclusions}
The main conclusion of this work is that CN remains in the
gas phase at densities close to $106$~\ccc\ in prestellar
cores. This result is surprising, given that CO, which is
the main repository of gas--phase carbon, is clearly
depleted at such densities. It follows that CN can serve as
a kinematic tracer of the high--density material. The
profiles of \nnhp\jone\ from A07 and of \thcn\ from the
present observations are in generally good agreement,
suggesting that both molecules trace the same density
regime. It is worth noting also that a double--Gaussian fit
to the \thcn\ spectrum in L~1544 at offset $(-20,20)$ yields
one component with a velocity of $7.300\pm0.011$~\kms\ and a
FWHM of $0.130\pm0.025$~\kms. The associated upper limit on
the kinetic temperature is $10.0\pm3.8$~K, which is in good
agreement with the value from the best fit temperature
profile of \cite{crapsi2007} at a distance of 30\arcsec,
where $\tkin=10$~K. In turn, this implies that the
non-thermal linewidth is at most 0.65 times the thermal
linewidth, and hence that turbulence has almost completely
dissipated.

CN is important also as a possible tracer of the magnetic
field in dense cores, by means of the Zeeman effect
\citep{crutcher1999}.  There exist already indirect
estimates of the magnetic field in L~1544 and L~183, based
upon the \cite{chandrasekhar1953bmag} method, which are of
the order of 100~$\mu$G \citep{crutcher2004a}. Independent
and direct measurements, based on the Zeeman effect, are
highly desirable. In this context, we note that OH Zeeman
measurements are sensitive only to the outer core, owing to
the large beam size and to the depletion of oxygen towards
the centre of the core.

The fact that CN is not depleted in the nuclei of L~1544 and
L~183 has interesting consequences for the chemistry in
these cores. One likely conclusion is that, although CO is
depleted by more than an order of magnitude at densities of
a few times $105$ to $106$~\ccc, there remains sufficient
CO in the gas phase to supply carbon to other, less abundant
species. Quantifying this statement will require model
calculations, which are under way. An expectation which is
based on the model calculations is that the effective $\rm
[C]/[O]$ gas--phase abundance ratio in dense depleted
regions, such as those in L~1544 and L~183, is larger than
the value of $\rm [C]/[O]=0.67$ adopted by
\cite{flower2005}.  The $\rm [CN]/[NO]$ abundance ratio is
sensitive to the amount of free oxygen (i.e. not tied up in
CO) in the gas phase. Our observations of L~1544 establish a
lower limit to $\rm [CN]/[NO]$ of order 0.1 in the high
density region; this suggests a $\rm [C]/[O]$ of at least
0.8. There will be consequences for HCN and HNC, and it
would be useful to determine limits on the abundances of
these species in the dense cores of L~1544 and L~183. We
note that our suggestion (A07) that CN acts as an
intermediary in the nitrogen chemistry is supported by the
present results.

\begin{acknowledgements}
We are indebted to the referee for some helpful and
perceptive comments on the original version of this Letter.
\end{acknowledgements}

\bibliographystyle{aa} \bibliography{general,cores,technic}

\begin{thebibliography}{26}
\expandafter\ifx\csname natexlab\endcsname\relax\def\natexlab#1{#1}\fi

\bibitem[{{Akyilmaz} {et~al.}(2007){Akyilmaz}, {Flower}, {Hily-Blant}, {Pineau
  des For{\^e}ts}, \& {Walmsley}}]{akyilmaz2007}
{Akyilmaz}, M., {Flower}, D.~R., {Hily-Blant}, P., {Pineau des For{\^e}ts}, G.,
  \& {Walmsley}, C.~M. 2007, \aap, 462, 221, (A07)

\bibitem[{{Belloche} {et~al.}(2002){Belloche}, {Andr{\'e}}, {Despois}, \&
  {Blinder}}]{belloche2002}
{Belloche}, A., {Andr{\'e}}, P., {Despois}, D., \& {Blinder}, S. 2002, \aap,
  393, 927

\bibitem[{{Bergin} {et~al.}(2002){Bergin}, {Alves}, {Huard}, \&
  {Lada}}]{bergin2002}
{Bergin}, E.~A., {Alves}, J., {Huard}, T., \& {Lada}, C.~J. 2002, \apjl, 570,
  L101

\bibitem[{{Bogey} {et~al.}(1984){Bogey}, {Demuynck}, \&
  {Destombes}}]{bogey1984}
{Bogey}, M., {Demuynck}, C., \& {Destombes}, J.~L. 1984, Can. J. Phys., 62,
  1248

\bibitem[{{Caselli} {et~al.}(1999){Caselli}, {Walmsley}, {Tafalla}, {Dore}, \&
  {Myers}}]{caselli1999}
{Caselli}, P., {Walmsley}, C.~M., {Tafalla}, M., {Dore}, L., \& {Myers}, P.~C.
  1999, \apjl, 523, L165

\bibitem[{{Chandrasekhar} \& {Fermi}(1953)}]{chandrasekhar1953bmag}
{Chandrasekhar}, S. \& {Fermi}, E. 1953, \apj, 118, 113

\bibitem[{{Crapsi} {et~al.}(2007){Crapsi}, {Caselli}, {Walmsley}, \&
  {Tafalla}}]{crapsi2007}
{Crapsi}, A., {Caselli}, P., {Walmsley}, M.~C., \& {Tafalla}, M. 2007, \aap,
  470, 221

\bibitem[{{Crutcher}(1999)}]{crutcher1999}
{Crutcher}, R.~M. 1999, \apj, 520, 706

\bibitem[{{Crutcher} {et~al.}(2004){Crutcher}, {Nutter}, {Ward-Thompson}, \&
  {Kirk}}]{crutcher2004a}
{Crutcher}, R.~M., {Nutter}, D.~J., {Ward-Thompson}, D., \& {Kirk}, J.~M. 2004,
  \apj, 600, 279

\bibitem[{{Dickens} {et~al.}(2000){Dickens}, {Irvine}, {Snell}, {Bergin},
  {Schloerb}, {Pratap}, \& {Miralles}}]{dickens2000}
{Dickens}, J.~E., {Irvine}, W.~M., {Snell}, R.~L., {et~al.} 2000, \apj, 542,
  870

\bibitem[{{Flower} {et~al.}(2005){Flower}, {Pineau des For{\^e}ts}, \&
  {Walmsley}}]{flower2005}
{Flower}, D.~R., {Pineau des For{\^e}ts}, G., \& {Walmsley}, C.~M. 2005, \aap,
  436, 933

\bibitem[{{Genzel}(1992)}]{genzel1992}
{Genzel}, R. 1992, in Saas-Fee Advanced Course 21: The Galactic Interstellar
  Medium, ed. W.~B. {Burton}, B.~G. {Elmegreen}, \& R.~{Genzel}, 275--391

\bibitem[{{Gerin} {et~al.}(1984){Gerin}, {Combes}, {Encrenaz}, {Linke},
  {Destombes}, \& {Demuynck}}]{gerin1984}
{Gerin}, M., {Combes}, F., {Encrenaz}, P., {et~al.} 1984, \aap, 136, L17

\bibitem[{{Hily-Blant} {et~al.}(2005){Hily-Blant}, {Pety}, \&
  {Guilloteau}}]{IRAM_report_2005-1}
{Hily-Blant}, P., {Pety}, J., \& {Guilloteau}, S. 2005, CLASS evolution: I.
  Improved OTF support, Tech. rep., IRAM

\bibitem[{{Leger} {et~al.}(1985){Leger}, {Jura}, \& {Omont}}]{leger1985}
{Leger}, A., {Jura}, M., \& {Omont}, A. 1985, \aap, 144, 147

\bibitem[{{Milam} {et~al.}(2005){Milam}, {Savage}, {Brewster}, {Ziurys}, \&
  {Wyckoff}}]{milam2005}
{Milam}, S.~N., {Savage}, C., {Brewster}, M.~A., {Ziurys}, L.~M., \& {Wyckoff},
  S. 2005, \apj, 634, 1126

\bibitem[{{{\"O}berg} {et~al.}(2005){{\"O}berg}, {van Broekhuizen}, {Fraser},
  {Bisschop}, {van Dishoeck}, \& {Schlemmer}}]{oberg2005}
{{\"O}berg}, K.~I., {van Broekhuizen}, F., {Fraser}, H.~J., {et~al.} 2005,
  \apjl, 621, L33

\bibitem[{{Pagani} {et~al.}(2007){Pagani}, {Bacmann}, {Cabrit}, \&
  {Vastel}}]{pagani2007}
{Pagani}, L., {Bacmann}, A., {Cabrit}, S., \& {Vastel}, C. 2007, \aap, 467, 179

\bibitem[{{Pagani} {et~al.}(2004){Pagani}, {Bacmann}, {Motte}, {Cambr{\'e}sy},
  {Fich}, {Lagache}, {Miville-Desch{\^e}nes}, {Pardo}, \&
  {Apponi}}]{pagani2004}
{Pagani}, L., {Bacmann}, A., {Motte}, F., {et~al.} 2004, \aap, 417, 605

\bibitem[{{Pety}(2005)}]{pety_gildas}
{Pety}, J. 2005, in SF2A-2005: Semaine de l'Astrophysique Francaise, ed.
  F.~{Casoli}, T.~{Contini}, J.~M. {Hameury}, \& L.~{Pagani}, 721

\bibitem[{{Pineau des For\^ets} {et~al.}(1990){Pineau des For\^ets}, {Roueff},
  \& {Flower}}]{pineau1990}
{Pineau des For\^ets}, G., {Roueff}, E., \& {Flower}, D.~R. 1990, \mnras, 244,
  668

\bibitem[{{Tafalla} {et~al.}(2002){Tafalla}, {Myers}, {Caselli}, {Walmsley}, \&
  {Comito}}]{tafalla2002}
{Tafalla}, M., {Myers}, P.~C., {Caselli}, P., {Walmsley}, C.~M., \& {Comito},
  C. 2002, \apj, 569, 815

\bibitem[{{Tafalla} {et~al.}(2006){Tafalla}, {Santiago-Garc{\'{\i}}a}, {Myers},
  {Caselli}, {Walmsley}, \& {Crapsi}}]{tafalla2006}
{Tafalla}, M., {Santiago-Garc{\'{\i}}a}, J., {Myers}, P.~C., {et~al.} 2006,
  \aap, 455, 577

\bibitem[{{van der Tak} {et~al.}(2005){van der Tak}, {Caselli}, \&
  {Ceccarelli}}]{vdtak2005}
{van der Tak}, F.~F.~S., {Caselli}, P., \& {Ceccarelli}, C. 2005, \aap, 439,
  195

\bibitem[{{Ward-Thompson} {et~al.}(1999){Ward-Thompson}, {Motte}, \&
  {Andr\'e}}]{ward1999}
{Ward-Thompson}, D., {Motte}, F., \& {Andr\'e}, P. 1999, \mnras, 305, 143

\bibitem[{{Williams} {et~al.}(1999){Williams}, {Myers}, {Wilner}, \& {di
  Francesco}}]{williams1999}
{Williams}, J.~P., {Myers}, P.~C., {Wilner}, D.~J., \& {di Francesco}, J. 1999,
  \apjl, 513, L61

\end{thebibliography}

\Online

\appendix
\section{Spectroscopic data}

Table~\ref{tab:spectro} gives the strengths and relative
intensities, $R$, of the 18 hyperfine components of the
$N=1-0$ transition of \th CN. The relative intensities are
such that the sum over all components is unity. The two
weakest hyperfine components [$F_1=1$, $(F_2\ra
F_2')=(2\ra1)$, $(F\ra F')=(1\ra2)$ and $(F_2\ra
F_2')=(0\ra1)$, $(F\ra F')=(1\ra0)$], which were omitted by
\cite{bogey1984}, are included here. We note that the
angular momentum coupling scheme is ${\bf S} + {\bf I}_1 =
{\bf F}_1$, ${\bf F}_1 + {\bf N} = {\bf F}_2$, and ${\bf
F}_2 + {\bf I}_2 = {\bf F}$, where $S = 1/2 = I_1$ and $I_2
= 1$; $S$ is the resultant electron spin quantum number, $N$
is the nuclear rotation quantum number, and $I_1$, $I_2$ are
the nuclear spin quantum numbers of $^{13}$C and $^{14}$N,
respectively (in the case of $^{12}$C, $I_1 = 0$).

\begin{table}
  \begin{center}
    \caption{Hyperfine structure components of the
      $(N'\rightarrow N)=1\rightarrow 0$ transitions in \th CN.
      $S$ is the line strength, and the relative intensities ($R$) are
      normalized to the sum of all the components. The
      selection rules are $\Delta F_1=0$, $\Delta
      F_2=0,\pm1$ and $\Delta F=0,\pm1$ \citep{bogey1984}.}
    \begin{tabular}{c c c c c c }
      \hline\hline
      $F_1$ & $F_2\ra F_2'$ & $F \ra F'$ & Frequency & $S(F,F')$ & $R$\\
      & & & MHz \\
      \hline
      0  &    1 \ra0   &     2\ra1    &  108651.2970  &   1.667  &  0.139\\
         &             &     1\ra1    &  108636.9230  &   1.000  &  0.083\\
         &             &     0\ra1    &  108631.1210  &   0.333  &  0.028\\
      1  &    2 \ra1   &     1\ra0    &  108786.9820  &   0.556  &  0.046\\
         &             &     2\ra1    &  108782.3740  &   1.250  &  0.104\\
         &             &     3\ra2    &  108780.2010  &   2.333  &  0.194\\
         &             &     1\ra1    &  108793.7530  &   0.417  &  0.035\\
         &             &     2\ra2    &  108796.4000  &   0.417  &  0.035\\
         &             &     1\ra2    &  108807.7879  &   0.028  &  0.002\\
      1  &    1 \ra1   &     1\ra0    &  108638.2120  &   0.333  &  0.028\\
         &             &     2\ra1    &  108643.5900  &   0.417  &  0.035\\
         &             &     1\ra1    &  108645.0640  &   0.250  &  0.021\\
         &             &     2\ra2    &  108657.6460  &   1.250  &  0.104\\
         &             &     0\ra1    &  108645.0640  &   0.333  &  0.028\\
         &             &     1\ra2    &  108658.9480  &   0.417  &  0.035\\
      1  &    0 \ra1   &     1\ra0    &  108406.0905  &   0.111  &  0.009\\
         &             &     1\ra1    &  108412.8620  &   0.333  &  0.028\\
         &             &     1\ra2    &  108426.8890  &   0.556  &  0.046\\
      \hline\hline
    \end{tabular}
    \label{tab:spectro}
  \end{center}
\end{table}

\section{Observed line ratios}
\def\ri{\ensuremath{\tilde{I}}}

The relative integrated intensities of the observed HFS
components have been computed at the $(0,0)$ positions in
both L~183 and L~1544 and compared with the values expected
in the optically thin limit, under the assumption of
LTE. These results are listed in Tables~\ref{tab:ratios12cn}
and \ref{tab:ratios13cn}.

\begin{table}[t]
  \begin{center}
    \caption{Integrated intensities of the
      CN$(N = 1\ra 0, F_2 = 3/2\ra 1/2)$ HFS components, relative
      to the $F = 5/2\rightarrow 3/2$ line, towards the
      (0,0) positions in L~183 and L~1544. The observed
      values of \ri\ are compared to the optically thin LTE
      prediction (Savage et al. 2002).}
    \begin{tabular}{c c c c c}\\
      \hline\hline
      $F\rightarrow F^{'}$ &  Frequency & \ri & \ri & LTE \\
      & MHz  &  L~183 & L~1544 \\\hline
      $1/2\rightarrow 3/2$ &  113520.414  &5.1$\pm$1.2 & 0.8$\pm$0.1 & 0.037  \\
      $3/2\rightarrow 3/2$ &  113508.934  &7.4$\pm$1.7 & 0.9$\pm$0.1 & 0.296  \\
      $1/2\rightarrow 1/2$ &  113499.643  &5.8$\pm$1.3 & 0.7$\pm$0.1 & 0.296  \\
      $5/2\rightarrow 3/2$ &  113490.985  &1           & 1           & 1 \\
      $3/2\rightarrow 1/2$ &  113488.142  &5.2$\pm$1.2 & 0.9$\pm$0.0 & 0.371 \\\hline
    \end{tabular}
    \label{tab:ratios12cn}
  \end{center}
\end{table}
\begin{table}[t]
  \begin{center}
    \caption{As Table~\ref{tab:ratios12cn} for the
      \thcn$(N = 1\ra 0, F_2 = 2\ra 1)$ HFS components,
      relative to the $F = 3\ra 2$ component.}
    \begin{tabular}{c c c c c}\\
      \hline\hline
      $F\rightarrow F^{'}$ &  Frequency & \ri & \ri & LTE \\
      & MHz  &  L~183 & L~1544 \\\hline
      $3\ra2$ &   108780.2010  &1             & 1             & 1      \\
      $2\ra1$ &   108782.3740  &0.40$\pm$0.10 & 0.65$\pm$0.15 & 0.536  \\
      $1\ra0$ &   108786.9820  &0.30$\pm$0.10 & 0.35$\pm$0.15 & 0.238  \\
      $1\ra1$ &   108793.7530  &0.25$\pm$0.12 &     $\le$0.40 & 0.179  \\
      $2\ra2$ &   108796.4000  &    $\le$0.10 &     $\le$0.30 & 0.179  \\\hline
    \end{tabular}
    \label{tab:ratios13cn}
  \end{center}
\end{table}

\section{Line profiles}

\begin{figure}
  \begin{center}
    \includegraphics[width=0.5\hsize,angle=-90]{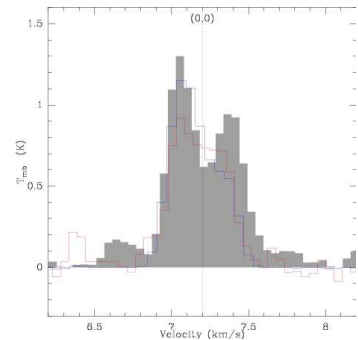}\\
    \caption{Comparison of line profiles towards L~1544 at
      offsets (0,0). CN: weakest HFS component at
      113520.4315~MHz (shaded histogram). \thcn: HFS
      component at 108780.2010~MHz, multiplied by 7 (red
      histogram). \nnhp: HFS component at 93176.2650~MHz,
      multiplied by 0.5 (blue histogram, from A07). The dashed
      line indicates the LSR velocity of 7.2~\kms.}
    \label{fig:profile_l1544}
  \end{center}
\end{figure}

\begin{figure}
  \begin{center}
    \includegraphics[width=0.5\hsize,angle=-90]{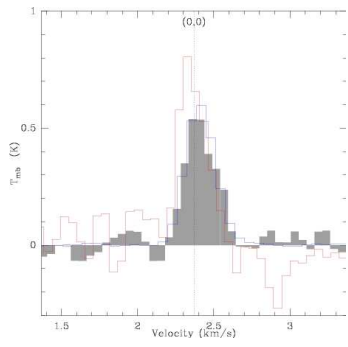}\\
    \caption{As Fig.~\ref{fig:profile_l1544}, but towards
      L~183. The HFS component at 93176.2650~MHz of \nnhp\
      is multiplied by 0.25 (blue histogram, from A07). The
      dashed line indicates the LSR velocity of
      2.37~\kms.}
    \label{fig:profile_l183}
  \end{center}
\end{figure}

Figs.~\ref{fig:profile_l1544} and \ref{fig:profile_l183}
show the profiles of the different tracers observed towards
L~1544 and L~183, respectively. The \nnhp\ data are taken
from A07. In L~1544, at those positions where the line is
double--peaked, all the profiles are mutually consistent,
except possibly at offset (-20,-20), where only the CN HFS
component shows two velocity components (but the
signal--to--noise ratio of the \thcn\ line is to low to draw
a definite conclusion). In L~183, all lines are consistent
at all positions.

\end{document}